# Revealing Hidden Connections in Recommendation Networks

**Rogério Minhano**
Universidade Federal do ABC - UFABC, Rua Santa Adélia., 166, Santo André, Brazil
discover.rogerio@tokiomarine.com.br

**Stenio Fernandes**
Universidade Federal de Pernambuco - UFPE, Av. Prof. Moraes Rêgo 1235, Cidade Universitária, Recife, Brazil
sflf@cin.ufpe.br

**Carlos Kamienski**
Universidade Federal do ABC - UFABC, Rua Santa Adélia., 166, Santo André, Brazil
cak@ufabc.edu.br

**Abstract:** Companies have been increasingly seeking new mechanisms for making their electronic marketing campaigns to become viral, thus obtaining a cascading recommendation effect similar to word-of-mouth. We analysed a dataset of a magazine publisher that uses email as the main marketing strategy and found out that networks emerging from those campaigns form a very sparse graph. We show that online social networks can be effectively used as a means to expand recommendation networks. Starting from a set of users, called seeders, we crawled Google's Orkut and collected about 20 million users and 80 million relationships. Next, we extended the original recommendation network by adding new edges using Orkut relationships that built a much denser network. Therefore, we advocate that online social networks are much more effective than email-based marketing campaigns..

**Categories:** J.4, I.6

## 1 Introduction

Viral marketing campaigns typically rely on word-of-mouth strategies, where existing users recommend products and services to their social networks. Explored by marketing professionals for decades, it is a well-known feature of human buying behaviour: people will be more interested in what a friend or acquaintance buys instead of randomly selecting a product. Those potential customers may buy the advertised product and/or send recommendations to a list of contacts who they believe they may have some influence on. This behaviour is usually promoted by rewarding customers with bonus products when a recommendation effectively is fulfilled with a purchase. Such strategies are called viral marketing, which means that the transmission of the advertising has an epidemic behaviour (i.e. spreading or increasing its occurrence like a disease). Viral marketing, in turn, is considered an instance of the more general idea of network-based marketing (Hill et al. 2006).

In the Internet age, companies have been increasingly using new media, such as email and text messages, to obtain a cascading recommendation effect similar to direct human contact. The first step in this direction is to build a large user database, with thousands or millions records, where each record may be of either an existing customer or a potential one. Depending on the objectives of each marketing campaign, the database is segmented in order to send targeted emails to the most promising subset of users. The following step is to provide incentives for users to propagate the advertising message, thus creating a recommendation network (RN). Some rough numbers say that typical return rates of such campaigns (in terms of purchases) are about 0.5%.

In this paper we analyse a database with subscription recommendations from a major Latin America magazine publisher using the theory of complex networks and their structural features, such as degree distributions, correlations among vertices, clustering coefficients, diameter, and average path lengths, in order to have a closer look at the viral marketing behaviour through network analysis. This database contains 28,562 people (acting as source and/or destination of recommendations) and 40,933 recommendations among them. We found out that, if we model the recommendations as a network, those people yield 9,562 sub-networks. In other words, those campaigns form a very sparse graph, where most sub-graphs have less than 3 vertices. This result shows that this type of campaign has a limited appeal among consumers, because most people do not follow the "recommendation chain". Most of the return rates of these campaigns come directly from massive number of emails sent to the user database, rather than to the emails that are propagated through user´s social network connections. Further analysis revealed that it is possible to classify users' behaviour in four well-defined types, namely "highly recommended people", "usual behaviour", "good recommenders", and "disseminators". Users are mapped into these classes according to their connections in the social network.

Our goal in this paper is to reveal the hidden relationships behind recommendation networks and to point out that Online Social Networks (OSN) can be effectively used as a means to promote viral marketing campaigns, by stimulating users to send recommendations of products or services to their friends. Starting from our existing dataset containing a sparse recommendation network, we crawled Google's Orkut and collected about 20 million users and 80 million connections among them. A set of subscribers were used as seeders (i.e., those who actively contribute to the social network community) for crawling the social network and as a result we built a large network of Orkut users, in a single large connected component. Although theoretically any other OSN might suffice for our purposes, Orkut was chosen because it is the preferred one in the location where this research was conducted. We searched thousands of users of the recommendation network manually in Orkut and found exactly 1,625 seeders, which in turn were used as the entry points in the Orkut network. This is itself a contribution of this paper, because we are not aware of another research work where approximately 20% of the current Orkut network have been collected and analysed using common metrics of complex networks. In the field of recommendation system, there are some research studies on how to use social graphs to boost the performance of recommendation systems (De Choudhury 2010) or to leverage viral marketing (Chen et al., 2010). However, our research work is different. Here recommendation network is also a social network and to the best of our knowledge, the idea of combining recommendation networks and social networks is new. Moreover, we dedicated great efforts to crawl and analyze two social network data. A full analysis on the extended network is provided, such as average path length, clustering coefficient, and the like.

Based on the relationships found in Orkut, we extended the original recommendation network, by adding new edges that built a much denser network. Those edges were given a weight in this extended recommendation network (ERN) according to the number of hops in social network, in order to make it possible to distinguish between them. It means that if two users of the RN are directly connected in Orkut, an edge is created in the ERN and assigned weight 1. If they are indirectly connected in Orkut through a single user, an edge is created in the ERN and assigned weight 2 and so on.

Results also indicate that there is a direct relationship between user behavior in the RN and Orkut. In both networks the seeders are among the most connected ones (i.e. a higher degree) and have friends highly connected too. In other words, an active user in terms of sending recommendations in the RN is also active in keeping a large list of friends in Orkut. This may have a significant impact for marketing strategies, since professionals may use this information to make better use of the effective recommenders in their campaigns. Therefore, we advocate that online social networks should be explorer by marketing campaigns.

This paper is organized as follows. Section 2 presents background information and discusses related work and section 3 explains the methodology used in this research. Sections 4 and 5 present our main results for the recommendation network and extended recommendation network respectively. We discuss the lessons learned and the possible outcomes from these results in section 6 and draw some conclusions in section 7

## 2 Background and Related Work

In this section, we present all the necessary technical background for an in-depth understanding of the paper. Also, we review the literature and show that our approach and results are unique and set the ground for further analysis of recommendation networks.

### 2.1 Viral Marketing

The study of epidemic behaviors in the network sciences area, like viruses spreading, and transmitting diseases, is highly relevant for understanding various areas that may be modeled as networks and their growing patterns (Kempe et al, 2003) (Iribarren et al, 2011) (Barash et al., 2012). A marketing technique called viral marketing has as its main feature the exploitation of this potential inherent to every social network. However, the path information travels to reach this epidemic stage is not straightforward. In some networks, the types of relationships are extremely important for a positive result. With the advent of the Internet, such advertising campaigns have been directed towards sending emails to potential customers, who may buy that product and/or send recommendations to a list of contacts who they believe they may have some influence on. The incentive companies use to promote this behavior is rewarding customers with bonus products.

The term viral marketing was first used in 1996 to describe the marketing strategy used by the free e-mail service Hotmail (Kaikati et al. 2004). Although this phenomenon has grown tremendously, there are still several interpretations for the term. Kiss and Bichler (Kiss and Bichler 2008), for example, define viral marketing as "marketing techniques that use social networks to produce increases in brand awareness by 'viral' diffusion processes, analogous to the spread of pathological and computer viruses." In other words, a company uses the social network of consumers as a way of popularizing a brand or product through the messages' dissemination. Bampo et al. (2008) define it as "a form of peer-to-peer communication in which individuals are encouraged to pass on promotional messages within their social networks". Viral marketing, in turn, is considered an instance of the more general idea of network-based marketing (Hill et al. 2006). Also, it is considered a type of word of mouth marketing, which aims at giving people reasons to exchange information about products/services and providing support for those conversations to take place (WOMMA 2005).

This strategy often works through electronic messaging (email) containing information about products and services. Phelps et al. (2004) suggest that "the forces driving the growth of email marketing are low costs to the marketer, the ability to target messages selectively, and high response rates relative to other forms of direct consumer contact." However, as the use of email marketing by businesses becomes more widespread, consumers are dealing with such messages as spam, increasingly diminishing rates of return in marketing campaigns. This factor is extremely important for understanding the success of viral marketing. Since emails from viral marketing strategies come from people one knows, consumers are much more reluctant to delete the message.

From the consumer's point of view, it is convenient to receive recommendations for products and services of interest. When searching for product information, people usually consult online blogs, communities, or the websites of vendors. According to Jupiter Research Institute studies, the majority of online shoppers use online tools to find interesting products (Loechner 2009). Although the study shows that the most popular search methods are search engines and e-commerce sites, 61% of consumers use recommendation messages as the basis for their purchases.

However, measuring the results of a viral marketing campaign is not trivial. According to Cruz and Fill (2008), only a limited number of research studies on the subject are available, which makes it impossible to determine what technique is most often used by professionals to measure such campaigns. They argue that it is very difficult to find a criterion to measure viral marketing because there are many ways in which users can be involved in a campaign. This interpretation is also supported by De Bruyn and Lilien (2008). They argue that it is difficult to explain why and how viral marketing works. Viral marketing campaigns result in peer-to-peer recommendations, thus increasing the credibility of the message. In addition, according to Rosen (2002) the acquisition of the product is part of a social process. This involves not only the interaction between company and customer, but also the exchange of information between people and the influences that are around the customer. The value of a customer for the company is not only related to the size of the purchase that he/she makes. His/her value should be measured by how many people on whom he/she can positively or negatively have an influence. According to Domingos (2005) well-connected consumers can help, but it is important that they like the product. Also, even though there are evidences that recommendations help people to make informed choices and therefore are considered a positive influence, sometimes the opposite happens (Fitzsimons and Lehmann 2004), e.g. when unsolicited advice contradicts someone´s initial impression.

A very important issue is how a member of a network is motivated to pass information. A person attending a cooking course will probably remember his/her classmates when he/she is involved in events related to cooking. In a bookstore, for example, if this person finds some books on issues raised during classes, he/she may consider that information being of great value to the group. At first, their relationships with classmates may just be based on the simple fact of sitting next to them while attending the cooking class. However, a new network is generated from the moment a book is shared within this group, which models attendees with a common interest on a particular subject. This type of network is similar to most common social networks, where communities sharing specific interests exchange information about products and events. When it comes to viral advertising campaigns, however, the motivation for having relationships are not necessarily of this type. In fact, the main purpose of a campaign is to capture the reason why someone passes on information that might be important to someone else. In the above example, a customer in a bookstore was the path to the book, to be known and possibly purchased by another person who may or may not be a book worm and knowledgeable of good publications. The main difference is that the impetus for the information disclosure was not caught. The publishers did not aim to generate situations wherein this behaviour occurs, or the customer was not stimulated by advertisements but by particular content.

However, in an advertising campaign via email, regardless of how good the product really is (and in fact the product quality always tends to be in the background in this type of disclosure) the campaign always tends to stimulate a momentary impulse and, as an exchange of favors, the stimulus tends to be better accepted if the participants are benefited. When recommendations are rewarded with a bonus, some people recommend dozens and become real good recommenders of advertising via email. On the other hand, others simply ignore the message and do not recommend to anyone. Also these campaigns are often targeted to a particular public, divided up by characteristics such as age, gender, income, and home address. This process generates an advertisement which is much more effective for that segmented public. People who develop such campaigns work with well-defined goals and a database containing details of potential customers. Therefore, advertising campaigns via email might no longer be considered a number of unwanted messages by potential customers, but might become highly productive mechanisms for both companies and customers. Its application must be carefully designed, since in a related scenario, i.e., in the Tweeter OSN, Harrigan et al. 2012 found that popular individuals have a significantly lower likelihood of *retweeting*, particularly when they are following a large number of individuals.

In this paper we aim at shedding some light into the advantages of the integration of viral marketing and Online Social Networks. By knowing their customer's social relationships, we expect marketing professionals will be able to create better and more effective campaigns.

## 2.2 Google's Orkut

Orkut is an OSN currently dominated by the Brazilian users, since 50.6% of its users come from Brazil. Google does not make available the number of Orkut users, but other non-official sources estimate this number to be over 100 million worldwide. Also, privacy policies of Orkut are not too strict when compared to other OSN and user profiles are public by default, which makes it appropriate for our purposes.

## 2.3 Complex Networks Theory

The analyses performed in this work are supported by the theory of complex networks, commonly used in social network analysis and based on graph theory. This subsection presents an overview of this theory. However, for an in-depth understanding of the subject we refer the user to the work of Newman (2003). Research on complex networks is multidisciplinary per se. Indeed it is closely related to disciplines such as physics, biology, mathematics, statistics, and computing. Most social networks have non-trivial characteristics with connections patterns between its elements that are neither regular nor random (Barabási and Albert 1999). Some characteristics include the degree distribution of vertices, the clustering coefficient, communities and hierarchies in such networks. The theory of complex networks has been widely used in the study of human interactions (Barabasi 2005). In the past decade, several research papers have been published in a number of areas, for example on the topological structure of the Internet (Pastor-Satorras et al. 2001), the World Wide Web (Broder 2000), online blogs (Leskovec et al. 2007), online social networks (Ahn et al. 2007), instant messaging networks (Leskovec and Horvitz 2008), scientific collaboration networks (Newman 2001), a network of sexual relations (Jones et al. 2003), prostitution networks (Rocha et al. 2010), and networks formed by geographical positioning (Liben-Nowell et al. 2005).

Traditionally, networks with complex topologies were described by the random graph model developed by Erdös and Rényi (1959). The ER model is simple because it assumes a fixed probability for a vertex to connect to another one, so that the resulting degree distribution of its vertices is Poisson. However, random graphs differ from reality as far as clustering of vertices and degree distributions are concerned, which led to the development of the small world and scale free networks models.

Unlike the ER model, vertices in real networks tend to be highly clustered. Also, the average distance between vertices is short even for large networks. Watts and Strogatz (1998) used the name small world networks to characterize networks that simultaneously present short distances and high clustering among their vertices.
Barabási and Albert (1999) created the concept of scale free networks, as they observed that in some real networks the probability of finding a highly connected vertex does not decrease exponentially as the vertex degree increases, as assumed by the ER model. Rather, they follow a power law distribution where their probability density function (PDF) has the form $p(x) \sim kx^{-\alpha}$, where $p(x)$ is the probability of finding the value x, k is a constant and $\alpha$ is known as the scaling parameter. In general, for most networks found in nature, the scale parameter lies between the limits two and three, i.e. $2<\alpha<3$ (Barabási and Bonabeau 2003). There are several ways to estimate the scale parameter of a power law distribution. A widely used approach is to construct a histogram of the data and chart it on a logarithmic scale of values. The result is a line very close to a straight line. However, in several cases, this method is not efficient and most of their results are not accurate as compared to more precise techniques such as Maximum Likelihood Estimation (MLE) (Clauset et al. 2009). In this study, we use the MLE technique.

## 2.4 Crawling Online Social Networks

It is well-known that access to user databases from OSN providers is not free, since there are privacy issues and business strategies involved. Therefore, we developed an automated collector (a crawler) for Orkut. Another important point is that collecting information of millions of OSN users is a computational challenge by itself. Many OSN services use dynamic resources in their Web pages (e.g. Ajax and DHTML) and complex types of interface customization. In addition, OSN servers have defense mechanisms against automated information collection, which raises the challenge to a new level of complexity.

Some OSN services, such as Facebook, Flickr, and Twitter, offer open libraries with APIs that help the process of data collection. However, Orkut only provides a library, called OpenSocial, aimed at application development that does not provide features for remote information gathering. Our crawler behaves as a web browser, selecting users´ friend lists, and treating the webpage for results cleaning up.

The choice of the seeders (RN users used as the entry points for crawling Orkut) is an important step in the process of collecting user information. Most examples in the literature use random seeders, which is an interesting approach when the goal is to capture a sample of the whole network. However there are some evidences that some features of the resulting graph are not equivalent to the real graph (Lee et al. 2006, Leskovec and Faloutsos 2006). In our work, the list of seeders is comprised of a set of users from the RN. Since our work focuses on finding out the relationship between the RN and another social network behind it, we prioritized the measurements around those seeders.

## 2.5 Related Work

Katona et al. (2011) address the effect of the network structure in diffusion of information in online social networks. They claim to be a structural pattern in the connections of the main influencers over the other participants of the network. Li et al. (2010) propose a framework for measuring the power of influence exerted by only reviewers over other people interested in a product. Their contributions show the importance of identifying the most influential vertices of a network. They propose that, after the identification of those vertices, companies should develop special marketing strategies for taking advantages of those reviewers. In both papers, the results confirm our hypothesis about the importance of finding out better ways of using the power of disseminators in recommendation networks. However our work has a different approach for dealing with disseminators in social networks. We demonstrate the existence of overlapping profiles with similar behaviors in two networks.

Leskovec et al. (2007) deal with networks of recommendations formed by viral marketing. They use a cascade model to analyze a network of 4 million people and 16 million recommendations. Both works deal with viral marketing, but from a different perspective (i.e. both use complementary approaches). Our work focuses on marketing and the work from Leskovec et al. focuses on sales. Our main interest is on marketing campaigns, particularly how to help marketing people in understanding how recommendation networks are formed and thus how new and more effective campaigns can be created. The work from Hill et al. (2006) surveys the area known as network-based marketing and presents data to provide empirical support those customers` social networks can affect the adoption of product or services. However, this work presents statistical models, but do not rely on network theory. In (Dinh, 20120), Dinh et al. explored the time aspect of influence propagation in OSN and demonstrated that viral marketing involves costly seeding, proportional to the size of the network. They concluded that the straightforward approach of targeting nodes with high degree in the network is no longer suitable when one needs the influence to propagate quickly throughout the network. Our work corroborates with their results through the analysis of complementary OSNs.

Swamynathan et al. (2008) study the impact over e-business of the connectivity of a user in social networks. They found out that transactions among friends or acquaintances in social networks result in a higher level of satisfaction for users. Their dataset comes from an e-commerce service that also behaves as a social network, which raises the question of how such services may use this information for their own benefit even though they do not have a social network. This is important because users are not willing to register in and keep their profiles up-to-date in a large number of social networks.

## 3  Methodology

In this section we describe the dataset and the methodology used for measurements and results analysis. Integration of the RN and the OSN was performed in four phases. Firstly, we describe the dataset of the original recommendation network (RN), where most users are connected to sparse subgraphs. Secondly, we show the process of Orkut crawling and measurement and the choice of the seeders. Thirdly, we deal with the extraction of a subgraph of the social network formed by the seeders, their friends, friends of their friends, and so on. Finally, we present how we integrated RN and OSN and the creation of a third network, called the extended RN (ERN).

### 3.1  Recommendation Data

The data analysed come from a database of subscription recommendations from a major magazine publisher in Latin America. We were able to collect data from the subscriber database where the analysed networks were stored. After that, we normalized them, i.e. we put them in a suitable format as recommendation network data. The dataset extracted is made of several sub-graphs, which in turn are composed of a very small number of nodes. More than 60% of the networks have 3 vertices or less and less than 15% of the networks have 10 vertices or more. By the time of extraction 28,562 people had recommended a product or received a recommendation and 40,933 recommendations occurred among these people. Those people yielded 9,562 sub-networks, the largest part with just a few recommendations. This analysis considered the period from February 2007 to December 2009. Figure 1 depicts a histogram of sub-network sizes.

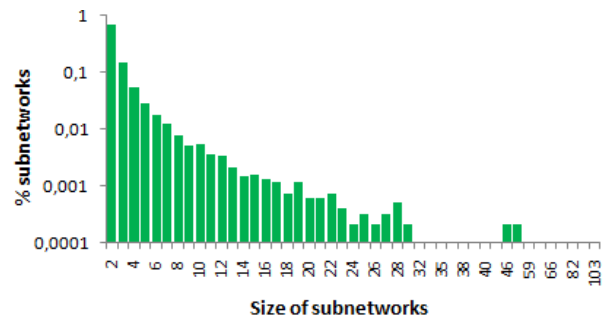

*Figure 1 - Sparse subgraph distribution in the recommendation network*

Although the most common network shapes found in the dataset are trees and stars, graphs of different forms arise from social relationships through which recommendations are propagated (Figure 2). These topologies suggest the existence of a real social network underneath the recommendation process.

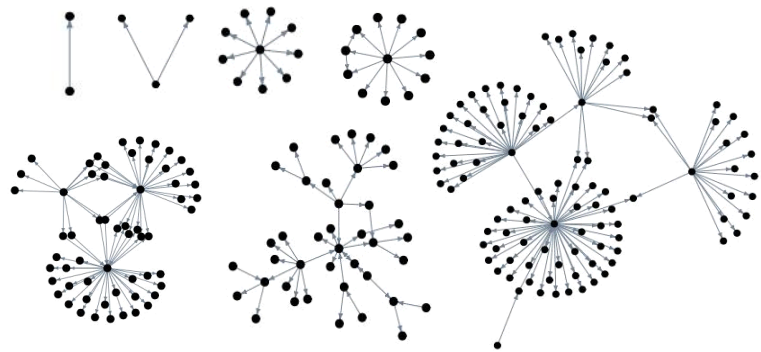

*Figure 2 - Typical topologies found in the recommendation network*

### 3.2  Finding Recommenders In Orkut

Out of 28,562 users in the RN we identified 5,600 ones who actually sent recommendations. Once we had those recommenders we went on by search Orkut after our potential seeders. Luckily (or purposefully, depending on the point of view) in Orkut users´ public profiles are open to all users by default, unlike Facebook, for instance, where you can only browse profiles of your friends. After that, we had a new challenge because neither user information in the RN is not particularly formatted for making a search easier nor Orkut has been planned to do so. Therefore, we manually "chased" RN recommenders in Orkut using a combination of different information available in each user´s profile, such as "name", "email", "location" and "age". The outcome was that we found 1,625 RN recommenders in Orkut and those are our seeders for the automated Orkut search.

Out of the four search filter options used, only "email" needs users´ authorization to appear in their profile. The other three options ("name", "location" and "age) are not affected by the standard information restriction policy. Therefore, for a significant number of users we had the three latter options only to find a user and decide whether they were the same person who made the recommendation. We only searched the 5,600 real recommenders in Orkut because only magazine subscribers can send recommendations, whereas for being recommended only an email address is needed (and this is the only information available). Since a great deal of Orkut users blocks access to their email addresses in their public profile, our choice was straightforward.

### 3.3 OSN Crawling Process

Once we had our list of seeders, we started the measurement crawling process to obtain the social graph, using a breadth-first search method. The measurement process is depicted in Figure 3 and it can be summarized by the following steps:

1. Create a node list initially with the seeder nodes;
2. Select the first seeder of the node list;
3. Collect its friend list;
4. Store its friend list in the end of the node list;
5. Remove the first node of the node list;
6. Go back to step 2 until all nodes have been visited.

In step 4 we stored the friend list into the node list up to the second level of the search. We observed that initially this approach generated a large number of sparse graphs with a tree shape. However, as the measurement process progressed, the number of edges became slowly higher than the number of vertices. The result of this process was that spontaneously the set of sparse graphs connected to each other up to a point where a single giant component was emerged. Numerical results from our measurements until the level 2 are presented in Table 1.

Table 1 – Graph sizes for levels 1 and 2

| Level | Seeders | Vertices | Edges |
|---|---|---|---|
| 1 | 1,625 | 244,686 | 252,726 |
| 2 | 244,686 | 21,697,657 | 79.136.104 |

The reasons we stopped the measurement in the second level are twofold. The first one is related to the number of Orkut users collected. Among all similar studies we found in the literature, our crawler collected the highest number of users. We do not have official data from Google, but this number is estimated to be over 100 million and therefore we collected around 20% of all Orkut users. For our purposes in the moment, which do not include to collect the entire Orkut network, this number is sufficient and significant. The second reason is related to the very goal of this work, which aims at creating a single component around the seeders for integrating the RN and the OSN. Therefore we considered that our needs have been fulfilled by this number of users.

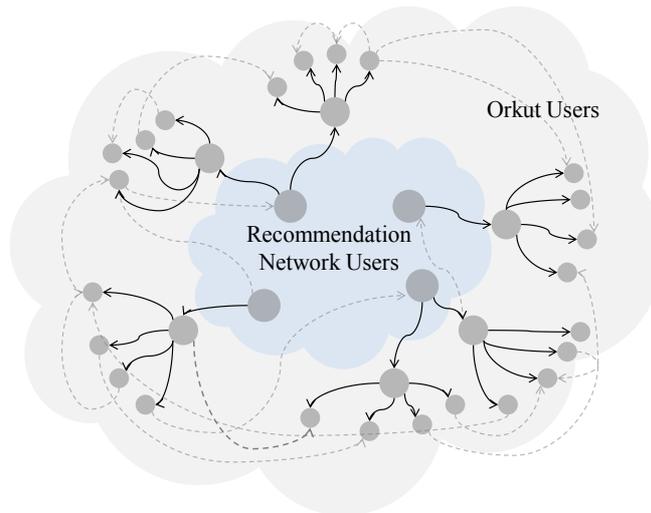

*Figure 3 - Finding Orkut users out of the RN seeder list*

### 3.4 Creating the Transition Network

After the measurement process in the Orkut network ended, we performed some transformations in the networks. Our first step was to extract a subnetwork (subgraph) using our seeders again as the starting point. This subnetwork is formed by the seeders and all Orkut users that appear in the shortest path among every pair of seeders. As a result we obtained a new network derived from the original Orkut network, which we called Transition Network (TN), with 1,072,785 vertices and 2,966,376 edges. This new component represents a relationship core among RN users inside the Orkut network. The TN is an appropriate tool for our purposes, since it is smaller than the Orkut network, thus making its analysis and further processing easier.

### 3.5 The Extended Recommendation Network

The last phase in the process of integrating the RN and the OSN is the creation of an Extended Recommendation Networks (ERN), which is the result of the connection of the various sparse subgraphs of the RN using the transition network. As shown in Figure 1, most subnetworks have no more than a couple of nodes. However, most subnetworks have topologies that suggest the existence of an underlying large social network. Knowledge of this network may have significant impacts for the marketing industry, both in the development of better campaigns and for providing a better understanding of who their customers actually are. We consider our ERN to be a prelude of such connections among users.

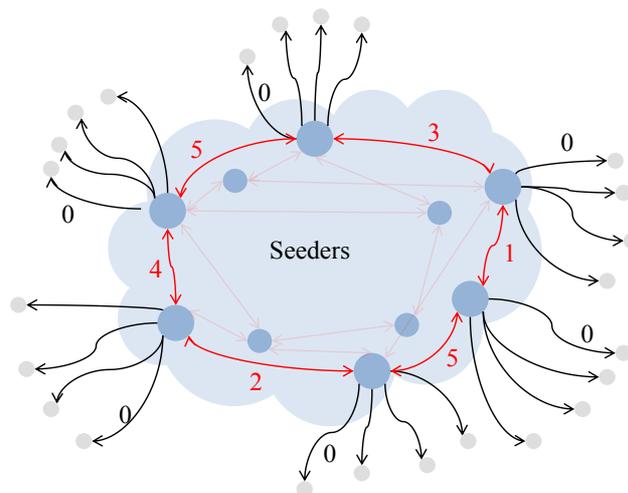

*Figure 4 – Combination between the RN and the TN to create the ERN*

The first step for the creation of the ERN is to compute the shortest path among all seeders in the TN, thus yielding a shortest path matrix. The next step is to combine the adjacency matrix of the RN with the shortest path matrix, creating new edges among seeders in the ERN. Each new edge is assigned a weight corresponding to the shortest path found in the transition network. We end up with a new network, denser than the original recommendation network, including the connections found for the seeders in Orkut, as depicted in Figure 4.

### 3.6 Metrics of Interest

We use common metrics for structural analysis of complex networks. There are good survey papers addressing how different metrics can be used in a number of contexts. We refer the reader to (Boccaletti et al. 2006, Costa et al. 2007) for an in-depth characterization of such metrics.

**Diameter**: the longest shortest path among all vertices of a graph. The choice of the longest shortest path is made by generating all shortest paths from each vertex to any other vertex in the network and then finally comparing them. At the end, the longest path is chosen.

**Average path length**: the average distance among all vertices of a graph. For this paper, it is the average of all shortest paths.

**Giant component**: the largest sub-graph contained in a disconnected graph. The recommendation database generated a large number of uncorrelated graphs and, thus, the largest among these graphs is the giant component.

**Degree distribution**: the degree of a vertex is the number of vertices adjacent to it. The distribution of degrees is commonly used in literature, and in this paper it is represented by the Empirical Cumulative Distribution Function (ECDF).

**Clustering coefficient**: Clustering coefficient (CC) is a metric that has particular importance on the theory of complex networks. It measures the tendency of some networks to form sets of grouped vertices. It is represented by the probability of finding a triangulation in a triple of vertices. In other words, assuming that vertex a is connected to vertices b and c, the CC is the probability that vertex b is connected to vertex c. Two types are commonly used, namely local and global CCs. A local CC indicates the percentage of clustering in a single vertex whereas the global one addresses the entire network (the average value of local CCs). In both cases, the result is a value between 0 and 1, which allows deriving insightful properties on the graph structures. A graph can be considered small world if its global CC is significantly higher than that of a random graph built on the same set of vertices, and both have roughly the same average shortest paths (Watts and Strogatz 1998). Here, CC implies that the more densely connected a subgraph is, the more likely a user will receive recommendations from several of his/her neighbors. This characteristic increases the likelihood of a user changing his/her mind.

**Asymmetry**: the degree of a vertex is the number of vertices adjacent to it. In directed networks, such as our recommendation network, the vertices have degrees of exit and entry. The ratio between these two metrics (input degree divided by the output degree) reveals important characteristics of how the interactions take place between network participants. In social networks, high correlation between vertices can be explained by the large number of symmetric edges. In other words, a network becomes highly symmetric when a user adds a friend to his/her list of friends and he/she is also added his/her friend's list.

**Assortativity**: this metric has the goal of showing the prevailing relationship between the vertices of the network. In this work, we evaluate assortativity by its visual representation, by observing the correlation between the average degree of the nearest neighbors of a vertex (knn) and the degree of the vertex itself (k). In other words, this metric yields a knn x k graph showing the frequency of relationships among vertices of different degrees. This may indicate that highly connected vertices tend to connect to less connected nodes or to vertices with degrees similar to theirs.

## 4 The Recommendation Network

### 4.1 Structure of the Recommendation Network

In order to get a better understanding of the recommendation network, we will analyze the two extremities of the dataset. Table 2 presents the frequency of sub-networks and their total number of vertices and edges (both indicated as percentages) contained in subnetworks with two to ten vertices. Subnetworks with more than ten vertices represent 3% of the total number whereas those with four vertices or less represent 90%. More interestingly, 70% of subnetworks have two vertices. In other words, the recommendation campaigns of our dataset usually do not generate large graphs, but a large number of small sub-graphs.

This information reveals the typical behavior of customers. Since a recommendation usually is rewarded with a bonus, some customers may just forward an email in order to obtain the bonus but effectively may not purchase anything. In other words, a large number of people with no special interest (or even worse, quite out of profile) receive the email, and the odds that the company will obtain new subscriptions decreases. Also, these people may not feel like forwarding the advertising email. This behavior actually dictates the effectiveness of an advertising campaign, which is greatly influenced by the size of the networks. Such campaigns seek to catch the buying impulse of people, purchasing products and also forwarding recommendations. This behavior is expected to yield a large recommendation chain.

Table 2 – Cumulative Distribution of Networks and Vertices

| Vertices (#) | Networks (%) | Vertices (%) | Edges (%) |
|---|---|---|---|
| 2 | 70,07 | 46,92 | 34,29 |
| 3 | 84,62 | 61,53 | 48,72 |
| 4 | 90,00 | 68,74 | 56,76 |
| 5 | 92,87 | 73,53 | 62,48 |
| 6 | 94,68 | 77,17 | 66,99 |
| 7 | 95,89 | 80,01 | 70,59 |
| 8 | 96,68 | 82,14 | 73,40 |
| 9 | 97,20 | 83,68 | 75,44 |
| 10 | 97,74 | 85,51 | 77,94 |

Table 3 – Dataset Analysis – Recommendation Network

| # Vertices | # Edges | Diameter | Avg. Path Length | Avg. CC | α |
|---|---|---|---|---|---|
| 103 | 222 | 5 | 1.766 | 0.000 | 1.52 |
| 92 | 236 | 3 | 1.009 | 0.014 | 1.20 |
| 82 | 179 | 3 | 1.000 | 0.000 | 1.25 |
| 77 | 180 | 2 | 1.000 | 0.000 | 1.15 |
| 66 | 130 | 4 | 1.211 | 0.000 | 1.46 |
| 59 | 156 | 4 | 2.441 | 0.000 | 1.99 |
| 46 | 261 | 3 | 1.078 | 0.007 | - |
| 40 | 103 | 6 | 1.838 | 0.024 | - |
| 33 | 74 | 6 | 1.162 | 0.000 | - |
| 26 | 63 | 4 | 1.037 | 0.018 | - |

Table 3 shows the ten largest and most representative components of the network. Last column shows the scaling parameter for all networks larger than 50 nodes (Clauste et al. 2009). The clustering coefficient (CC) is the probability that for a given graph whose vertex *a* is connected to a vertex *b* and whose vertex *b* is connected to a vertex *c*, its vertex *a* is also connected to the vertex *c*. CC is zero for 60% of the subgraphs analyzed. In addition, the diameter (the longest of the shortest paths among all vertices) is also low. Therefore, we may conclude that these networks demonstrate that some customers are very efficient in their recommendations. Furthermore, 40% of the components have the CC higher than zero, even in small diameters. For all sub-graphs, the number of edges is higher than the number of vertices, which in some cases reaches up to three times the former by the latter. This leads us to believe that communities formed between existing customers maintain a continuous exchange of recommendations based on existing social networks, for example friends, relatives, acquaintances, and the like.

### 4.2    Symmetric Behavior of Recommendations

In order to understand symmetries in behaviors between recommendations sent and received we calculated the ratio between outdegree and indegree for all vertices in all sub-networks. We found out four different groups of users according to their behavior in a viral marketing campaign, as depicted by the CDF in Figure 5:

**Highly recommended people - $p(x \leq 0.1)$**: potential customers, i.e. people who receive many recommendations but are not that productive at recommending, either because they do not like the product or consider received emails as spam. Since they receive many recommendation messages, their social network is working well, at least inbound. In our network they count for 10% of the users.

**Usual (symmetric) behavior - $p(0.1 < x \leq 0.75)$**: people who send and receive the same number of recommendations on average. We consider this behavior as usual because around 65% of the recommendations are exchanged by reciprocal relationships between users. A sound hypothesis for explaining this behavior is the pursuit of getting the reward. Another possibility is just that most people prefer to follow a moderate middle path, between highly recommended people and disseminators.

**Good recommenders - $p(0.75 < x \leq 0.9)$**: people who usually send more recommendations than they receive at a rate of 2-4 times, count for 15% of our user base.

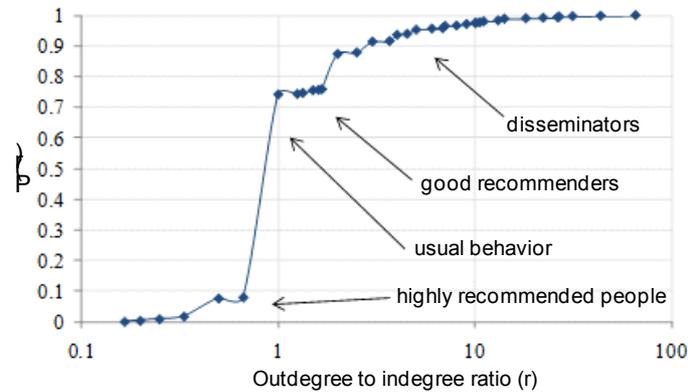

*Figure 5 – Outdegree to indegree ratio reveals groups of users with different behaviors*

**Disseminators - $p(0.9 < x)$**: people who behave as disseminators, i.e., they are prolific at sending recommendations but do not receive many of them. People who have a disseminator profile correspond to only 10% of our user base.

As far as symmetry in recommendations is concerned, we observed that 65% of the relationships are symmetric, whereas 35% are asymmetric. In other words, there is an evident symmetrical behavior of most users that may be used to direct the focus of marketing professionals during the development of new campaigns.

### 4.3 Assortativity in Recommendations

Assortativity is usually applied to relationship networks revealing the preference of users to connect to others who are similar or not. We calculated this metric for all our subnetworks, since our largest component alone is not significant. In Figure 6 we can observe the assortativity of the recommendation network, i.e. the relationship between a node degree (*k*) and the average degree of its neighbors (*knn*). There are two complementary findings in this picture. First, the higher the degree of a node, the lower the average degree of its neighbors. Second, the lower the degree of a node, the higher the average degree of its neighbors. This pattern of relationship indicates that the largest networks are composed of highly connected users who exchange recommendations with lowly connected users. This finding implies that the largest components are generated by disseminators. Although they are 10% of our user base, their behavior is much more relevant for the evolution of a subnetwork than the behavior of the other users. Therefore, the more effective a disseminator is, the greater is the likelihood of expansion of the larger components. In other words, effective disseminators significantly contribute to the creation of larger subnetworks, since they send many recommendations (i.e. create new edges) to a number of users (either add new vertices or connect existing subnetworks).

There are only a few cases where both *k* and *knn* have high values, which show that very active users, as far as sending recommendations is concerned, rarely establish relationships with each other. We conjecture here that the generation of larger components in email-based marketing campaigns necessarily involves the creation of specific campaigns where disseminators are stimulated to send recommendations to each other, even though they may not know each other a priori. This strategy will induce the connection of disconnected subnetworks into larger components and from this point on new relationships among highly connected users can be leveraged in such a way as to achieve more effective campaigns, by stimulating the epidemic behavior among users.

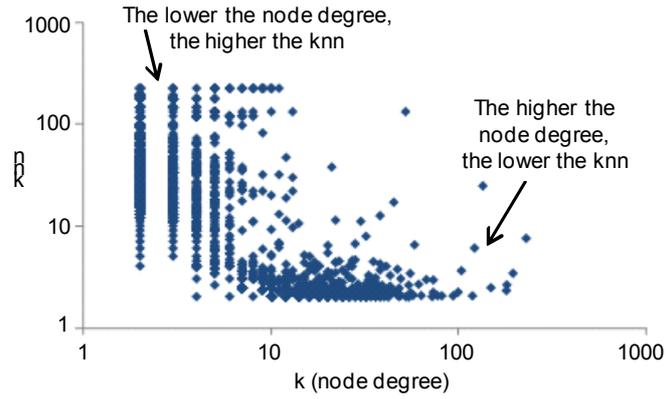

*Figure 6 – Users´ preference when connecting to each other*

## 5 Extending the Recommendation Network

### 5.1 Structural properties of Orkut and Transition Network

Table 4 shows the results for the measurement process in Orkut and the extraction of the transition network. The Orkut network forms a single giant connected component with about 21 million vertices. As far as we know, this is the largest subgraph of the Orkut network ever presented in the literature.

The clustering coefficient is similar to the results presented in other studies (Clauste et al. 2009), as well as the scale parameter (α) that indicates a power law distribution. However, the average number of friends of each Orkut user is lower than the one found in Clauste et al.( 2009). We consider that this is due to the bias created by our measurement method. As we performed a breadth-first search, in the last level there will be always a large number of leaf nodes, i.e. vertices with only one edge. Those low-degree vertices have a significant impact on decreasing the average degree. On the other hand, the average degree of the seeders is much higher than the general average, since they are in the core of our network. Since we collected two complete levels starting with the seeders, they and their friends faithfully correspond to a significant part of the whole relationship network in Orkut.

This behavior can also be found in the TN, where the difference between seeders and the all set of nodes is even higher (more than 14 times). Also, the clustering coefficient is considerably lower for the transition network compared to our whole Orkut network. Even though the transition network is a subset of the Orkut network, its properties are affected by the extraction mechanism. We extracted only the nodes that belong to a given shortest path between two seeders. All other nodes were discarded. In addition, there may be more than one shortest path connecting two seeders, from which we had to choose only one. Therefore, intermediate nodes from different shortest paths may not be coincident, which means that we made no effort to build a minimal transition network, since for our purposes this was unnecessary. The result is that the average degree and the clustering coefficient of the transition network were affected by this strategy.

Table 4 – Orkut and Transition Network

| Network | Vertices | Edges | CC | α | l | Avg. Degree | Avg. degree seeders |
|---|---|---|---|---|---|---|---|
| Orkut | 21M | 79M | 0.18 | 1.78 | 4.82 | 72 | 318 |
| Transition | 1M | 3M | 0.0025 | 3.83 | 4.13 | 10 | 144 |

As for the average path length (l) it becomes relatively stable before and after the extraction of the transition network, because the network was formed around the seeders and a single component was build so that it is quite obvious that the paths will be short. Also, since the seeders and their intermediate fellows were preserved in the transition network the average path length is only slightly lower than our Orkut network.

**Degree Distribution**

The node degree distribution for various complex networks of different types follows a power-law (section 2.3). Our network shows a characteristic consistent with a power-law, i.e. a small number of vertices with a large number of edges and a large number of vertices with a small number of edges. In Figure 7 we show the degree distributions for Orkut and the TN. We calculated the scale parameter of the power-law for both networks. For Orkut we actually calculated three different values for the scale parameter. The first one for the nodes with degrees between 1 and 70, the second one for degrees between 71 and 800 and third one is the general parameter that is presented in Table 4. In the tail of the distribution, marked in the figure as a vertical dashed line, are the users with more than 800 friends. The figure shows that they are part of a small minority of users and there is an artificial limit in Orkut where each user can have no more than 999 friends. Very popular users go around this limitation by creating new profiles. However, the search and combination of those profiles into a single one may be a slow and painful manual job, which is out of the scope of our work.

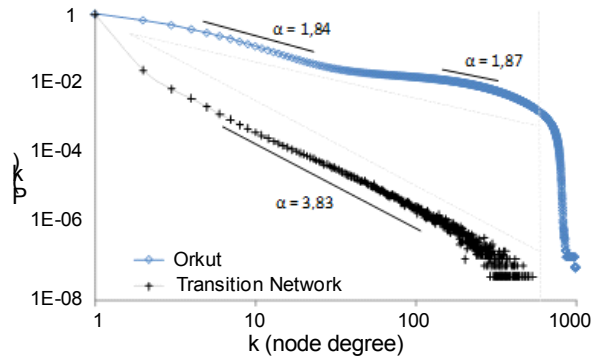

*Figure 7 – Degree distribution for Orkut and Transition network*

The distribution for the transition network has a more distinguished behavior. Even though the value 3.83 is relatively high for this type of network, it has almost the shape of a straight line. This means that the probability that we choose by chance a vertex with more than 800 edges is much lower than the probability of choosing a vertex with a smaller number of edges. In the center of the picture we drew two dashed lines that form a triangle (with the vertical line). They show that the degree of the more well connected nodes decreased during the process of extracting the transition network. It happens because even though a user has a large number of friends (i.e. a high degree), only the nodes which are intermediate ones in a given shortest path were transferred to the transition network. Also, we may observe that no user with more than 800 friends included in the TN.

**Clustering Coefficient and Degree Correlation**

Figure 8 presents the clustering coefficient (CC) as a function of the node degree, for the Orkut and transition networks. In both networks the CC tends to decrease as the degree increases. Orkut presents an increase for users with 50 to 100 friends, which suggests the presence of a highly connected local cluster. The average CC among seeders is 0.099 which is close to this zone and is considered an indication that a core has been created around the seeders. We also may observe the same behavior among the users with more than 800 friends. This suggests that those users are also part of this connected core.

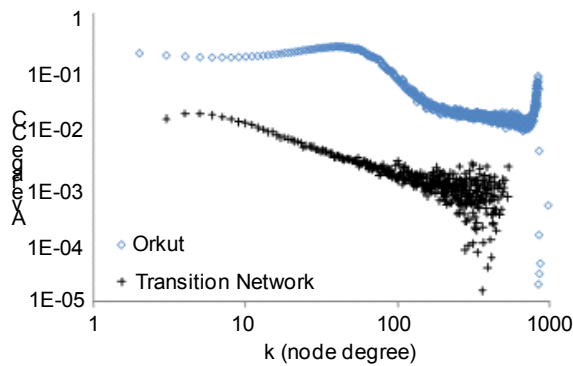

*Figure 8 – Clustering Coefficient as a function of the node degree*

Degree correlation is a way of getting and understanding of user relationship patterns in a network. Figure 9 shows the average degree of the neighbors (knn) of a node as a function of the node degree (k). We can observe that users with many friends tend to be connected to users with fewer friends and vice-versa for both networks. However, in our Orkut sample network there exist a clear change (decrease) for users between 50 and 100 friends, which coincides with the increase of the curve pointed out in Figure 8. The average degree of the neighbors of the seeders is 86, whereas the degree of the seeders is 318. This information corroborates with the existence of a tightly connected core around the seeders. Also, in both Figure 8 and Figure 9 there is an increase in the curve for nodes with a degree higher than 700, which happens because those users are connected to the seeder-based core.

The TN present values of clustering coefficient and degree correlation lower than the Orkut network, which can be explained by the same reason pointed out at the analyses of power-law distributions (in this section). The extraction process only selects the nodes that appear in the intermediate paths among the seeders within only one possible shortest path. All in all, it follows the same behavior of Orkut.

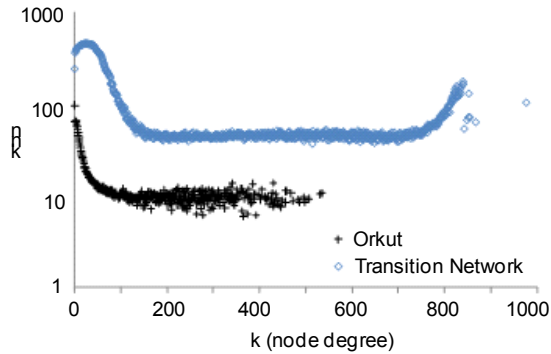

*Figure 9 – Assortativity for Orkut and Transition networks average degree of the neighbors (knn) vs. node degree (k)*

## 5.2 The seeders inside the networks

Figure 10 presents the empirical cumulative probability distribution (ECDF) for seeders´ degrees in the recommendation network (left) and for the network sizes where seeders belong to (right). Among all seeders, the most and least connected ones have degrees 21 and 2 respectively. Only 11% of the vertices in the recommendation networks have degrees higher or equal to 2, implying that seeders are among the most active recommenders in the network.

The sizes of their networks show the same behavior, i.e., they belong to the larger (sparse) networks. Thus, they are active users and so are their friends. Actually, it was no surprise that those seeders were found in Orkut, since they are active in other social networks too. In other words, users have a typical and well defined "social" profile and they tend to keep it in different social networks.

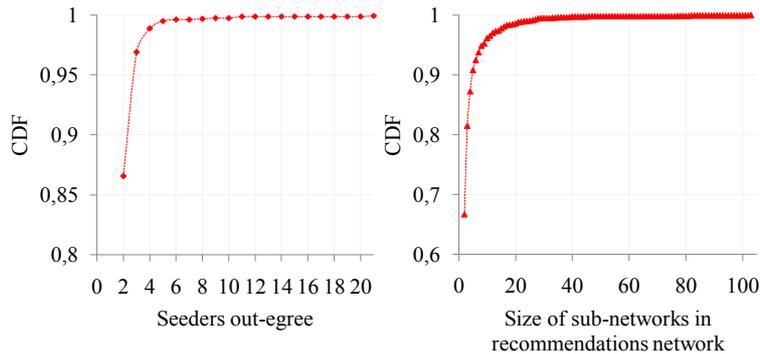

*Figure 10 – ECDF for seeders´ degrees and network sizes in the RN*

Figure 11 shows a comparison between the degree distribution for the seeders and all users in our Orkut network. A quite significant portion of Orkut users collected in our measurement (about 90%) has less than 10 friends. As mentioned before, this is an effect of the measurement process based on a breadth-first search, which tends to create a highly connected core around the seeders. However, there are only 1,625 seeders whereas the remaining 10% users count 2 million. The average number of friends among our sample Orkut network is 72 and for the seeders this value is 318, which obviously means that seeders are more connected than the average user of our Orkut network.

Results presented in this section demonstrate the strong relationship between the behavior of the seeders in the recommendation and Orkut networks. In both they are among the most connected users and they are also friends of the most connected ones. This suggests that different profiles of the same user in different networks are overlapping, i.e. in a email recommendation network and in Orkut.

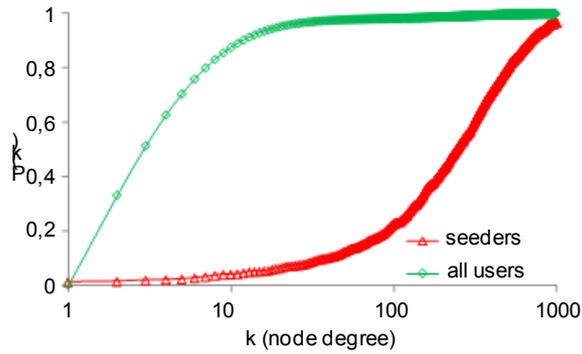

*Figure 11 – ECDF for the seeders and all users in Orkut*

This finding (which is quite intuitive but not easy to prove) brings significant positive repercussions for marketing professionals, because new strategies may be used to make a better use of the potential social profile of the recommenders.

### 5.3 Extended Recommendation Network

The Extended Recommendation Network (ERN) is build by the combination of the Recommendation Network (RN) by adding new relationships (edges) that come from our chosen Online Social Network (OSN), i.e. Orkut. The process of building the ERN maintains the same number of vertices ofthe original RN, with extra edges. The goal is to create a larger connected component (the largest component in the RN has 103 nodes) and therefore being able to support our claim that the use of social networks for viral marketing may have a better result than simply using email messages.

Based on the relationships found in Orkut, we extended the original recommendation network, by adding new edges that built a much denser network. Those edges were given a weight in this extended recommendation network (ERN) according to the number of hops in social network, in order to make it possible to distinguish between them. It means that if 2 users of the RN are directly connected in Orkut they are assigned weight 1. If they are indirectly connected in Orkut through a single user, an edge is created in the ERN and assigned weight 2 and so on. In the sequence, we will focus our analysis in the largest component in the ERN by progressively adding progressively new edges.

Table 5 summarizes the main structural characteristics of those new large components compared the original RN, where we refer to each network by the maximum weight of Orkut-originated included in the analysis. Therefore, five ERNs will be analyzed, where the first one (ERN1) includes only the new Orkut-originated edges with weight 1 (we call them 1-weighed edges) and the last one (ERN5) includes edges with weights 1, 2, 3, 4 and 5. The columns called Weight x show the number of edges with weight x added to network ERNx (where ERN0 = RN). We can observe that the original RN did not receive any other weighted edge. On the other hand, ERN1 has 117 edges, 106 from the RN (i.e. weight 0) plus 11 with 1-weighted edges. Also, ERN2 has 6691 edges, 3253 from the RN (it combined more sparse subnets than ERN1) plus 35 1-weighted edges and 3425 2-weighted edges.

Table 5 – Main Structural Characteristics of the Extended Recommendation Network

| Net | # Vert. | # Edges | Avg. Path Length | CC | Diam. | Weight 0 | Weight 1 | Weight 2 | Weight 3 | Weight 4 | Weight 5 |
|---|---|---|---|---|---|---|---|---|---|---|---|
| RN | 103 | 106 | 1.76 | 0.000 | 5 | 106 | - | - | - | - | - |
| ERN1 | 103 | 117 | 3.03 | 0.000 | 5 | 106 | 11 | - | - | - | - |
| ERN2 | 4480 | 6691 | 6.39 | 0.052 | 31 | 3253 | 35 | 3403 | - | - | - |
| ERN3 | 5381 | 184115 | 3.39 | 0.336 | 14 | 3868 | 35 | 3415 | 176797 | - | - |
| ERN4 | 5395 | 1109039 | 2.63 | 0.918 | 13 | 3876 | 35 | 3415 | 176797 | 924916 | - |
| ERN5 | 5395 | 1267112 | 2.50 | 0.982 | 12 | 3876 | 35 | 3415 | 176797 | 924916 | 158073 |

The small number of 1-weighted edges shows that only a small number of seeders are directly connected in Orkut. This happens because marketing campaigns usually have a nationwide scope and most users do not know each other directly. However, when it comes to 2-weighted edges and over the impact on marketing strategies of using an OSN becomes evident. The highest impact in the network aggregation process is made by 2- and 3-weighted edges. In other words, with as little as 2 or 3 hops in the Orkut network we were able to aggregate the majority of the interesting recommendation sparse subnetworks. These edges significantly contribute to the creation of a single giant component inside de recommendation network. Comparing the results for ERN2 with Orkut, we can observe that the average of the shortest paths is higher (6.39), whereas ERN3 is similar to Orkut.

Table also reveals some emerging patterns for network structural properties. For a better illustration of these patterns, we calculated the relative percentages (where 100% is the largest value) that the values of those properties represent for all ERNs. Figure 12 shows that the more edges we add in the RN, the lower the diameter and the average path length are. It is worth emphasizing that for calculating the shortest paths edge weights are considered, so that if a given shortest path crosses a 2-weighted edge, 2 hops are summed up. Therefore, the shortest paths for ERN4 and ERN5 shows that both networks became highly connected. ERN4 is a thorough argument that with no more than four hops any user can be reached in the network, among those 5,395 users that can actually be contacted from each other. This means that in our experiment more than 23,000 users could neither be connected to each other, nor to the giant connected component. We think that those users either do not use social networks, or they could be reached by a deeper search into Orkut (adding more nodes), or even by searching other social networks or other socialization services (examples are instant messaging or VOIP systems). Figure 12 clearly shows that with as little as three hops we can reach practically 100% of the vertices, but two hops are more cost-effective (80%). This is because there is a social "cost" associated with going beyond any number of hops. For example, when a user receives a marketing message, they may decide whether to recommend it further to their social network or not. Obviously that the longest the radius a message is propagated in a social network, the highest its probability to die. Therefore, since our goal is to connect the highest number of people though the recommendation network, we consider ERN3 as the ideal situation.

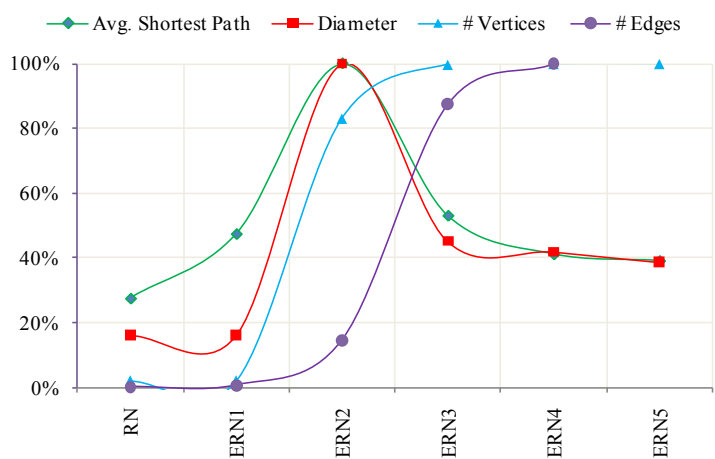

*Figure 12 – Behavior of some ERN structural properties*

Figure 13 shows that the progressive addition of new edges has the effect of making node degree tend to a uniform distribution, i.e. the probability of randomly choosing a node with a low or high degree becomes the same, which is really uncommon in real world networks. The scale parameter (α) also reveals this transformation, varying from 2.74 in ERN1 to 2.03, 1.46 and 1.31 in ERN2, ERN3 and ERN4 respectively. Please note that networks following a power law usually present the scale parameter in the range between 2 and 3 (section 2.3).

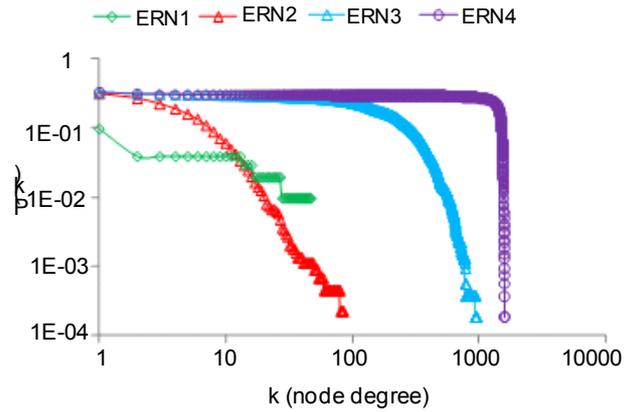

*Figure 13 – Degree distribution for the Extended Recommendation Networks (ERN)*

The clustering coefficient (CC) of the extended networks also brings important information. For the RN and ERN1, CC is around zero, because most subnetworks have a tree shape, i.e. there is almost no transitivity among nodes. As more edges are added, it grows to 0.336 for ERN3 and almost 1 for ERN5. This information may have a strong impact on marketing campaigns, since the same message may arrive from different sources to a user, and its critical threshold for acceptance may be overpassed more easily (Centola 2010). Figure 14 depicts the average CC of the extended networks as a function of the node degree. Once more it reveals that the addition of edges coming from Orkut changes the network structure. ERN3 presents the common behavior of the CC decreasing as the node degree increase (Mislove et al. 2007). However, for ERN4 (and ERN5, not shown in the picture) the CC is close to 1 to most node degrees.

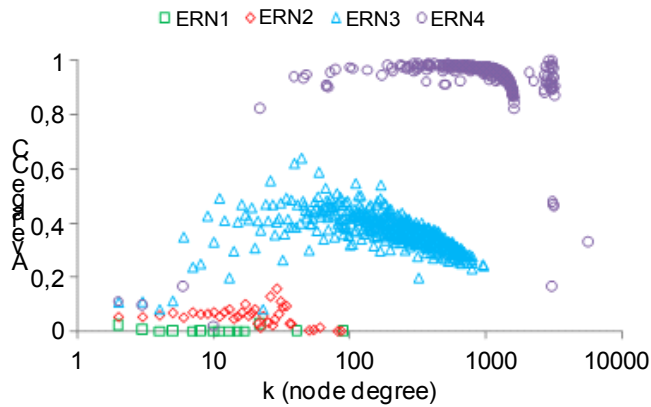

*Figure 14 – Average Clustering Coefficient (CC) for the Extended Recommendation Networks (ERN)*

In a complete graph, where all vertices are connected to each other, the average degree of the neighbors of a node (the knn) is N-1 where N is the number of vertices. The assortativity of the extended networks, in Figure 15, clearly shows that vertices with high degree are connected to each other, which comes from the relatively uniform increase in the number of edges for most vertices in the network. From ERN3 to ERN4 only a few new vertices are added to the largest component, whereas around 900,000 new edges connected the existing vertices. In other words, with as little as 3 hops in the OSN, most users become highly connected nodes in the recommendation network.

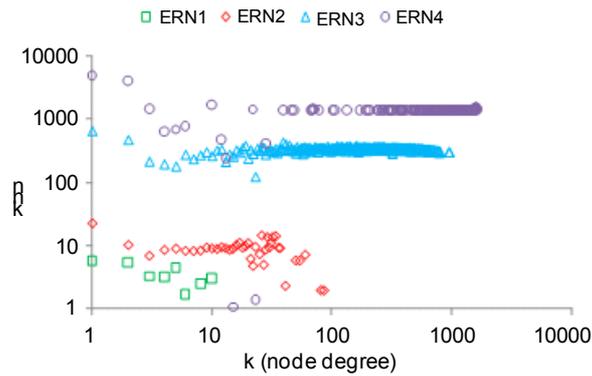

*Figure 15 – Average degree of the neighbors (knn) vs. node degree (k) for the Extended Recommendation Networks (ERN)*

### 5.4 Degree distribution for the ideal ERN

ERNs analyzed in section 5.3 are recommendation networks where we added new weighted edges. This transformation causes huge changes in the structural properties of the network (e.g., clustering coefficient, average degree and average path length). In order to understand the real consequences for our ideal network (ERN3), we added the intermediate nodes (and their edges) from Orkut into ERN3. As shown in Figure 16 the scale parameter for the degree distribution for this new network is 1.69, which indicates that its behavior is also coherent with a power-law distribution. The main outcome is that if degree distribution follows a power-law this network may be characterized as a scale-free network, thus having highly connected nodes (called hubs) that make the network robust and have a fundamental role in transmitting information through the network (Newman 2003).

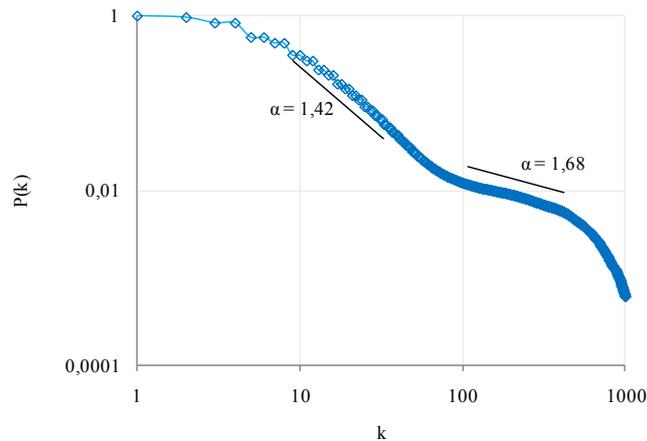

*Figure 16 – Degree distribution for ERN3*

### 5.5 Shortest paths made even shorter

Our last results specifically address the impacts of the ERN in the shortest path lengths calculated only among seeders. Figure 17 shows that for Orkut this value is always close to five, which is coherent with the average shown in Table .

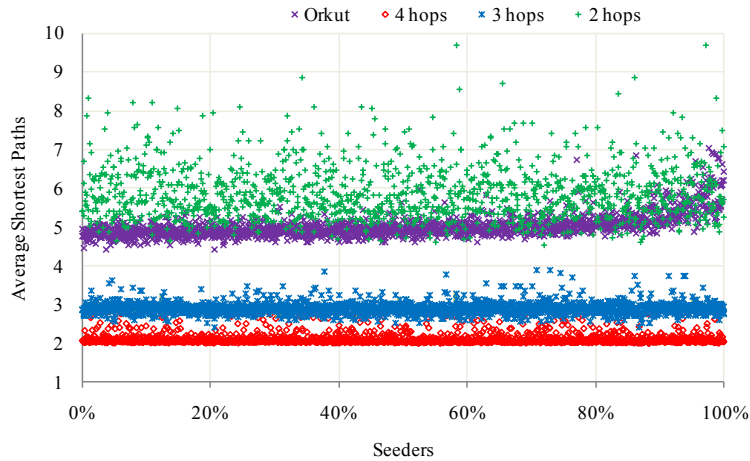

*Figure 17 – Average shortest path length among seeders*

For ERN2, however, this is not intuitive in a first analysis, since the lengths are higher than for Orkut. For ERN3 the situation changes again. What happens is that in the original RN the paths are always short because most networks are disconnected. When we added some new edges, we connect a large number of previously unconnected networks and therefore path lengths increase in a first moment. If we continue adding edges, the number of new networks (and nodes) added decrease and the number of shortcuts increase. Therefore, path lengths tend to decrease, but this effect is only reached for ERN3 and ERN4, where the values are 3.39 and 2.63, respectively.

These results make us believe that the shortest paths become even shorter when we combine two networks. Another observation that emerged from the dataset is that relationships among people in the Internet do not happen only through specific relationship sites. Rather, in various OSNs users have different profiles and may adopt a different behavior. As a matter of fact, the integration of networks gives us the opportunity to know different profiles and behaviors of the same user.

## 6 Discussion

### 6.1 Email-based viral marketing vs. epidemic behavior

An important metric for analyzing the success of a viral marketing campaign is whether it can convince a large number of people to work for it, e.g. by sending recommendations to their contacts. Companies use mechanisms to send emails that contain marketing activities for thousands of users. Only a small percentage of users forward a given message as an endorsement of the product to their friends. As a result, a relevant question can be raised on this approach: Can it be even considered as viral marketing? It is expected that a viral marketing campaign has an epidemic behavior. From the data related to the size of the network and to the number of vertices, we observed that epidemic behavior is rarely achieved. This conclusion suggests that the success of a campaign is related to the size of direct email sent. It is not related to the potentiality of the marketing strategy to turn a campaign into a viral one. In other words, success depends more on the number of emails sent than on user acceptance. However, results mentioned above show that the formed networks appear to have characteristics similar to real social networks (for example, scale-free networks), where people send recommendations to their friends for convenience, need, or kindness.

Also we could identify that recommendation networks can be separated into four distinct groups: highly recommended (10%), usual behavior (65%), good recommenders (15%) and disseminators (10%). From these figures, "usual" cases predominate. However, in fact, the exceptions here represent thousands of people. Therefore, it is strongly recommended that marketing campaigns should be broken down into groups so that one can properly handle such exceptions. Analysis of disseminators confirms this hypothesis, since the major components (subnetworks) always tend to be generated by them. Therefore, we suggest that a successful strategy to create viral marketing campaigns should be done by establishing a single social network in which all persons interested in the products are connected. From the results of this work, we can infer that the most direct way of achieving this is through stimulus and incentives to disseminators. Thus, we speculate that the creation of strategies based on social relationships can make a marketing campaign to achieve a viral behavior. Therefore, stimulating the recommendations between separate networks by creating targeted campaigns is a very important step for existing recommendation networks to become connected.

### 6.2 Using Online Social Networks for viral marketing

Currently, many strategies for the development, advertisement and sale of new products and services are based on marketing campaigns. However, common techniques used for advertising those campaigns do not seem to have followed the evolution of the online social networks. Many companies use classic data mining techniques, such as segmentation by gender, age, and income in their campaigns. On the other hand, networks formed by marketing campaigns are social networks too, even though they connect people who share particular interests. Therefore, a direct contribution of our work is to reveal the existence of social relationships behind recommendation networks.

We found out that the most active users in the recommendation network, the disseminators, play the same role in Orkut, where they have a high number of friends. The very existence of this relationship involving users of the two networks reveals a social behavior in the first place. Disseminators are not only active when they may obtain a reward. Rather, they are active in social networks in a broad sense. Therefore, marketing professionals must know the social relationships of users who have a disseminator behavior pattern, in order to improve their strategies. Since most disseminators in the our recommendation network were found to be up to 3 hops away from each other in Orkut, marketing campaigns aiming at putting them in contact may have a much significant impact in terms of recommendations exchanged. Also, this may be yield interesting behaviors in terms of achieving the main goal of viral marketing, i.e. increasing sales or information dissemination.

## 7 Conclusion

This paper analyzed the integration of two social networks, which are a magazine recommendation network and an online social network, namely Orkut. This integration yielded a third network, called extended recommendation network, a combination of various disconnected subgraphs from the recommendation network with the seeder transition network from Orkut.

We collected Orkut users from a group of initial users extracted from the recommendation network, called seeders. This process created a single connected component around the seeders with 21 million vertices and 80 million edges. Our results show that seeders tend to be among the most connected and also to be friends with the most connected in both networks. As far as we know, this work is the first to present real evidences of this relationship. We showed that only 3 hops in a social network may have a huge impact in the creation of an extended recommendation network.

This work opened up avenues for further investigation. For example, as future work, we intend to explore with more details the integration between the recommendation and online social networks. Also, we think that other online social networks, as well as other relationship services in the Internet may also help the creation of effective recommendation networks.